\documentclass[12pt]{iopart}
\usepackage{iopams}
\expandafter\let\csname equation*\endcsname\relax
\expandafter\let\csname endequation*\endcsname\relax
\usepackage{amsmath}
 
\begin{document}

\title{Potentials and fields of a charge set suddenly from rest into uniform motion}
\author{V Hnizdo$^1$ and G Vaman$^2$} 
\address{$^1$ 2044 Georgian Lane, Morgantown, WV 26508, USA}
\address{$^2$ Aleea Callatis 1, Bucharest, Romania}
\eads{\mailto{hnizdo2044@gmail.com} and \mailto{getavaman@gmail.com}}
\begin{abstract}
The fact that electromagnetic effects propagate at the speed of light suggests how the Lorenz-gauge scalar and vector potentials of a uniformly moving point charge must be modified when the charge was initially at rest and then set suddenly into uniform motion. The modified potentials are shown to satisfy the requisite inhomogeneous wave equations. The gauge function of the transformation of these potentials to the Coulomb gauge is calculated in closed form. It is validated by confirming that the Coulomb-gauge vector potential that is calculated using it yields together with the Coulomb-gauge scalar potential the same electric and magnetic fields as those calculated with the Lorenz-gauge potentials.
\end{abstract}
\noindent{\it Keywords\/}: classical electrodynamics, Lorenz gauge, Coulomb gauge, gauge function
\section{Introduction}
A point charge $q$ moving with a constant velocity ${\bi v}=v\hat{\bi x}$ along the $x$-axis so that it passes through the origin ${\bi r}=0$ at a time $t=0$ generates electric and magnetic fields
\begin{align}
{\bi E}({\bi r},t)=q\frac{(x-vt)\,\hat{\bi x}+y\,\hat{\bi y}+z\,\hat{\bi z}}{\gamma^2[(x-vt)^2+(y^2+z^2)/\gamma^2]^{3/2}},\quad {\bi B}({\bi r},t)= \frac{\bi v}{c}\times{\bi E}({\bi r},t).
\label{EB}
\end{align}
Here and henceforth $\gamma=(1-v^2/c^2)^{-1/2}$; the Gaussian units are used. These are the well-known fields of a uniformly moving charge, first found by Oliver Heaviside, already some 135 years ago \cite{Heavi}.While the corresponding Lorenz-gauge scalar and vector potentials,
\begin{align} 
\Phi_{\rm L}({\bi r},t)=\frac{q}{\sqrt{(x-vt)^2+(y^2+z^2)/\gamma^2}},\quad {\bi A}_{\rm L}({\bi r},t)=\frac{v}{c}\,
\Phi_{\rm L}({\bi r},t)\,\hat{\bi x},
\label{PA}
\end{align}
are well known also, the corresponding Coulomb-gauge vector potential,
\begin{align}
A_{{\rm C}x}({\bi r},t)&=A_{{\rm L}x}({\bi r},t)+\frac{c}{v}\,[\Phi_{\rm C}({\bi r},t)-\Phi_{\rm L}({\bi r},t)],
\label{Ax}\\
A_{{\rm C}y}({\bi r},t)&=-\frac{c}{v}\,\frac{y(x-vt)}{y^2+z^2}[\Phi_{\rm C}({\bi r},t)-\Phi_{\rm L}({\bi r},t)],
\label{Ay}\\
A_{{\rm C}z}({\bi r},t)&=-\frac{c}{v}\,\frac{z(x-vt)}{y^2+z^2}[\Phi_{\rm C}({\bi r},t)-\Phi_{\rm L}({\bi r},t)],
\label{Az}
\end{align}
where 
\begin{align}
\Phi_{\rm C}({\bi r},t)=\frac{q}{\sqrt{(x-vt)^2+y^2+z^2}}
\label{Phic}
\end{align}
is the `instantaneous' Coulomb-gauge scalar potential, is not known so well. In \cite{VH1}, it was obtained by calculating the gauge function of the transformation from the Lorenz gauge to the Coulomb gauge, refuting a claim \cite{On1} that the electric field of a uniformly moving charge comes out differently when it is calculated in the Coulomb gauge.

What are the fields, and the potentials, of a point charge that has {\it not} been moving uniformly at all times, but was set in such motion from an initial state of rest?   In his classic text \cite{Pur}, Purcell uses the fact that electromagnetic effects propagate at the speed of light to modify the electric field of equation\,(\ref{EB}) accordingly, under  the simplifying assumption that the onset of the charge's motion occurs `suddenly'.\footnote{A similar problem, namely that of the fields a uniformly moving charge that is stopped suddenly, was dealt with in a similar way already by J J Thomson \cite{Thom}.} In this paper, we shall treat Purcell's model in proper mathematical terms, starting with the Lorenz-gauge potentials.  Our results will  confirm his informal findings.  We shall also calculate the gauge function of the transformation of the Lorenz-gauge potentials of the initially resting charge to the Coulomb gauge, and, using it, obtain the Coulomb-gauge vector potential that replaces the vector potential (\ref{Ax}--\ref{Az}) under the changed kinematics of the charge's motion.

\section{Lorenz-gauge potentials} 
\noindent The charge and current densities of a point charge $q$ that had been at rest at the origin ${\bi r}=0$ until time $t=0$ and then it started  suddenly to move with a constant velocity $v$ along the $x$-axis are given by
\begin{align}
\tilde{\rho}({\bi r},t)= q\,\delta[x -vt\Theta(t)]\,\delta(y)\,\delta(z),\quad
\tilde{\bi J}({\bi r},t)&= vq\,\delta(x-vt)\delta(y)\delta(z)\Theta(t)\hat{\bi x}.
\label{tilderhoJ} 
\end{align} 
Here, $\Theta(\cdot)$ is the Heaviside step function, which we shall treat, on a par with the Dirac delta 
function, as a generalized function; thus $\Theta(0)$ is undefined and $\rmd\Theta(x)/\rmd x= \delta(x)$.
Despite the unphysical jump of the charge's velocity from zero to a nonzero value, the continuity equation is satisfied. Writing the charge density of equation\,(\ref{tilderhoJ}) as 
\begin{align}
\tilde{\rho}({\bi r},t)=q\delta(x-vt)\delta(y)\delta(z)\Theta(t)+q\delta(x)\delta(y)\delta(z)\Theta(-t)
\label{tilderho2}
\end{align}
we have, indeed, using the identity $\delta(x -vt)\delta(t)=\delta(x)\delta(t)$  and  the expression of equation\,(\ref{tilderhoJ}) for $\tilde{\bi J}$,  
\begin{align}
\partial\tilde{\rho}({\bi r},t)/\partial t&=-v\delta'(x-vt)\delta(y)\delta(z)\Theta(t)+\delta(x-vt)\delta(y)\delta(z)\delta(t)- \delta(x)\delta(y)\delta(z)\delta(t)\nonumber \\
&=-v\delta'(x-vt)\delta(y)\delta(z)\Theta(t)\nonumber \\
&=-\boldsymbol{\nabla}\bdot\tilde{\bi J}({\bi r},t).
\label{continuity}
\end{align}

The Lorenz-gauge potentials, as the electromagnetic field itself, are governed by wave equations according to which electromagnetic effects propagate at the speed of light. Following Purcell \cite{Pur}, we use this fact to modify the Lorenz-gauge potentials (\ref{PA}) of a uniformly moving charge to those generated by the charge and current densities (\ref{tilderhoJ}), as follows:
\begin{align}
\tilde{\Phi}_{\rm L}({\bi r},t)=\Phi_{\rm L}({\bi r},t)\,\Theta(ct-r)+(q/r)\,\Theta(r-ct),\;\;
\tilde{\bi A}_{\rm L}({\bi r},t)={\bi A}_{\rm L}({\bi r},t)\,\Theta(ct-r).
\label{tildePA}
\end{align}
It can be shown that the potentials (\ref{tildePA}) satisfy the Lorenz-gauge condition
\begin{align}
\boldsymbol{\nabla}\bdot\tilde{\bf A}_{\rm L} +\partial\tilde{\Phi}_{\rm L}/c\partial t =0.
\label{Lcon}
\end{align}
Indeed, 
\begin{align}
\boldsymbol{\nabla}\bdot\tilde{\bf A}_{\rm L}&=\beta\frac{\partial \Phi_{\rm L}}{\partial x}\,
\Theta(ct-r)-\beta\frac{x}{r}\Phi_{\rm L}\,\delta(ct-r),\\
\frac{\partial\tilde{\Phi}_{\rm L}}{c\partial t}&=-\beta\frac{\partial\Phi_{\rm L}}{\partial x}\,
\Theta(ct-r)+(\Phi_{\rm L}-q/r)\,\delta(r-ct),
\end{align}
where here and henceforth $\beta=v/c$, and so
\begin{align}
\boldsymbol{\nabla}\bdot\tilde{\bf A}_{\rm L}+\partial\tilde{\Phi}_{\rm L}/c\partial t &=
\left[(1-\beta x/r)\Phi_{\rm L}-q/r\right]\delta(ct-r)\nonumber \\
&=\left[(1-\beta x/r)\Phi_{\rm L}|_{t=r/c}-q/r\right]\delta(ct-r)=0.
\end{align}
The last line here obtains because 
\begin{align}
(1-\beta x/r)\Phi_{\rm L}|_{t=r/c}-q/r &=q(1-\beta x/r)/\sqrt{(x-\beta r)^2+(r^2-x^2)(1-\beta^2)}-q/r\nonumber \\
&=q(1-\beta x/r)/|r-\beta x| -q/r=0.
\end{align}

If the potentials (\ref{tildePA}) satisfy the requisite inhomogeneous equations,
\begin{align}
\Box\,\tilde{\Phi}_{\rm L}({\bi r},t)=-4\pi\tilde{\rho}({\bi r},t),\quad
\Box\,\tilde{\bi A}_{\rm L}({\bi r},t)= -(4\pi/c)\tilde{\bi J}({\bi r},t),
\label{weqs}
\end{align}
where $\Box=\nabla^2-\partial^2/c^2\partial t^2$ is the d'Alembertian operator, then these potentials 
are indeed the Lorenz-gauge potentials of a point charge set suddenly from rest into uniform motion. 
The calculation  of  $\Box\tilde{\Phi}_{\rm L}$ is facilitated by the use of the identity
\begin{align}
\Box [f({\bi r},t)g({\bi r},t)] = (\Box f)g+f\Box g+2[\boldsymbol{\nabla}f\bdot\boldsymbol{\nabla} g-(\partial f/c\partial t)(\partial g/c\partial t)].
\label{Box}
\end{align}
Thus
\begin{align}
\Box\,\tilde{\Phi}_{\rm L}({\bi r},t)=&(\Box \Phi_{\rm L})\Theta(ct{-}r)+\Phi_{\rm L}\Box\Theta(ct{-}r)\nonumber \\
&+2[\boldsymbol{\nabla}\Phi_{\rm L}\bdot\boldsymbol{\nabla}\Theta(ct{-}r)-(\partial \Phi_{\rm L}/c\partial t)
(\partial \Theta(ct{-}r)/c\partial t)]\nonumber \\
&+[\Box(q/r)]\Theta(r{-}ct)+(q/r)\Box \Theta(r{-}ct)+2 \boldsymbol{\nabla}
(q/r)\bdot \boldsymbol{\nabla}\Theta(r{-}ct).
\label{BoxtildeP}
\end{align}
To evaluate (\ref{BoxtildeP}), we need these results
\begin{align}
\Box\Phi_{\rm L} &=-4\pi q\delta(x-vt)\delta(y)\delta(z),\label{BoxP}\\
\Box\Theta(ct-r)&=-(2/r)\delta(ct-r),\\
\boldsymbol{\nabla}\Phi_{\rm L}\bdot\boldsymbol{\nabla}\Theta(ct-r)
&=\frac{q\gamma\delta(ct-r)[\gamma^2 x(x-vt)+y^2+z^2]}{r[\gamma^2(x-vt)^2+y^2+z^2]^{3/2}},\\
\frac{\partial \Phi_{\rm L}}{c\partial t}&=\frac{q\beta\gamma^3 (x-vt)}{[\gamma^2(x-vt)^2+y^2+z^2]^{3/2}},
\label{dPhiLdt}\\
\frac{\partial \Theta(ct{-}r)}{c\partial t}&=\delta(ct-r),\\
\Box(q/r)&=-4\pi q\delta(x)\delta(y)\delta(z),\\
\Box\Theta(r{-}ct)&=(2/r)\delta(r-ct),\\
\boldsymbol{\nabla}(q/r)\bdot \boldsymbol{\nabla}\Theta(r{-}ct)&=-(q/r^2)\delta(r-ct).
\end{align}
Here, the result (\ref{BoxP}) expresses the fact that the charge density that produces the Lorenz-gauge scalar  potential of equation\,(\ref{PA}) is that of a uniformly moving charge. With these results, we obtain 
\begin{align}
\Box\,\tilde{\Phi}_{\rm L}({\bi r},t)=&-4\pi q\big[\delta(x{-}vt)\delta(y)\delta(z)\Theta(ct{-}r)
+\delta(x)\delta(y)\delta(z)\Theta(r{-}ct)\big]\nonumber \\
&+q\frac{2\beta\gamma^3(x{-}vt)(ct{-}r)\delta(ct{-}r)}{r[\gamma^2(x-vt)^2+y^2+z^2]^{3/2}} 
\nonumber \\
=&-4\pi q\big[\delta(x-vt)\delta(y)\delta(z)\Theta(ct-|vt|)+\delta(x)\delta(y)\delta(z)\Theta(-ct)\big]\nonumber \\
=&-4\pi \tilde{\rho}({\bi r},t),
\label{BoxtildeP2}
\end{align}
where the delta-function identities  $f({\bi r})\delta({\bi r}-{\bi a})=f({\bi a})\delta({\bi r}-{\bi a})$ and
$x\delta(x)=0$ are used to obtain the third line. The  scalar potential $\tilde{\Phi}_{\rm L}$ is thus shown to
satisfy the requisite inhomogeneous wave equation. It can be shown similarly that also the vector potential 
$\tilde{\bi A}_{\rm L}$ satisfies the requisite inhomogeneous wave equation, $\Box\,\tilde{\bi A}_{\rm L}=-(4\pi/c) \tilde{\bi J}_{\rm L}$.

\section{Fields} 
Using the Lorenz-gauge potentials (\ref{tildePA}), we can now calculate the electric  field of a point charge set suddenly from rest into uniform motion. This field is given by
\begin{align}
\tilde{\bi E}({\bi r},t)&=-\boldsymbol{\nabla}\tilde{\Phi}_{\rm L}({\bi r},t)-\partial \tilde{\bi A}_{\rm L}({\bi r},t)/c \partial t
\nonumber \\
&= -(\boldsymbol{\nabla}\Phi_{\rm L})\Theta(ct-r)-\Phi_{\rm L}\,\boldsymbol{\nabla}\Theta(ct-r) +(q\hat{\bi r}/r^2)\,\Theta(r-ct)
-(q/r)\boldsymbol{\nabla}\Theta(r-ct)\nonumber \\
&\,\,\,\,\, -\beta(\partial\Phi_{\rm L}/c\partial t)\Theta\,(ct-r)\hat{\bi x}-\beta\Phi_{\rm L}\,\delta(ct-r)\hat{\bi x}\nonumber \\
&= -(\boldsymbol{\nabla}\Phi_{\rm L})\Theta(ct-r)+\Phi_{\rm L}\,\hat{\bi r}\,\delta(r-ct) +(q\hat{\bi r}/r^2)\,\Theta(r-ct)
-(q\hat{\bi r}/r)\delta(r-ct)\nonumber \\
&\,\,\,\,\, +\beta^2(\partial\Phi_{\rm L}/\partial x)\Theta(ct-r)\hat{\bi x}-\beta\Phi_{\rm L}\,\delta(ct-r)\hat{\bi x}.
\label{tildeE}
\end{align}
First, we collect and examine the delta-function terms,
\begin{align}
[(\hat{\bi r}-\beta\hat{\bi x})\,\Phi_{\rm L}|_{t=r/c} -q\hat{\bi r}/r]\delta(ct-r)&=q\left( \frac{\hat{\bi r}-\beta\hat{\bi x}}{r-\beta x}-\frac{\hat{\bi r}}{r}\right)\delta(r-ct)\nonumber \\
&=q\frac{\beta(x\hat{\bi r}-r\hat{\bi x})}{r(r-\beta x)}\,\delta(r-ct).
\label{delta}
\end{align}
This non-vanishing  delta-function term can contribute to the electric field only on the spherical surface 
$r=ct$. The full electric field (\ref{tildeE}) is thus 
\begin{align}
\tilde{\bi E}({\bi r},t)=&-\left[(1-\beta^2)\frac{\partial\Phi_{\rm L}}{\partial x}\,\hat{\bi x}+\frac{\partial\Phi_{\rm L}}{\partial y}\,\hat{\bi y}+\frac{\partial\Phi_{\rm L}}{\partial z}\,\hat{\bi z}\right]\Theta(ct{-}r)\nonumber\\
&+
\frac{q\hat{\bi r}}{r^2}\,\Theta(r{-}ct)+q\frac{\beta(x\hat{\bi r}-r\hat{\bi x})}{r(r-\beta x)}\,\delta(r{-}ct).
\label{fulltilde}
\end{align}
In a final form, this reads
\begin{align}
\tilde{\bi E}({\bi r},t)={\bi E}({\bi r},t)\,\Theta(ct-r)+\frac{q\hat{\bi r}}{r^2}\Theta(r-ct)
+q\frac{\beta(x\hat{\bi r}-r\hat{\bi x})}{r(r-\beta x)}\,\delta(r-ct),
\label{finaltildeE}
\end{align}
where ${\bi E}$ is the Heaviside electric field of equation\,(\ref{EB}).
For times $t>0$, within a sphere of radius $r=ct$, the electric field (\ref{finaltildeE}) reduces to the Heaviside field, outside this sphere, it is the static Coulomb field $q\hat{\bi r}/r^2$ of a point charge located on the origin $r=0$; on the infinitesimally thin spherical surface $r=ct$ itself, the field is, strictly speaking,  not defined. For times $t<0$, the field equals $q\hat{\bi r}/r^2$ at all values of $r$. These results confirm the informal findings in the classic  text of Purcell \cite{Pur} for the field of a `charge that starts'. 

The Lorenz-gauge vector potential of equation\,(\ref{tildePA}) produces the magnetic field corresponding to the electric field (\ref{finaltildeE}):
\begin{align}
\tilde{\bi B}({\bi r},t)&=\boldsymbol{\nabla}\times\tilde{\bi A}_{\rm L}({\bi r},t)\nonumber \\
&=\boldsymbol{\beta}\times {\bi E}({\bi r},t)\,\Theta(ct-r)+q\frac{\boldsymbol{\beta}\times\hat{\bi r}}{r-\beta x}\,\delta(r-ct),
\label{tildeB}
\end{align}
where ${\bi E}$ is again the Heaviside electric field; the delta-function term arises through the factor $\Theta(ct-r)$ in the vector potential. Within the sphere of radius $r=ct$, this magnetic field equals the Heaviside magnetic field of a uniformly moving charge, but outside this sphere, the magnetic field vanishes.

The electromagnetic field $(\tilde{\bi E}, \tilde{\bi B})$ of equations\,(\ref{finaltildeE}) and (\ref{tildeB}) was obtained using the Lorenz-gauge potentials (\ref{tildePA}) 
that satisfy the requisite inhomogeneous wave equations,
$\Box\tilde{\Phi}_{\rm L}=-4\pi\tilde{\rho}$ and $\Box\tilde{\bi A}_{\rm L}=-(4\pi/c)\tilde{\bi J}$, where 
the charge density $\tilde{\rho}$ and the current density $\tilde{\bi J}$ are those of a point charge set instantaneously from rest into uniform motion but still satisfying  the continuity equation. In view of these facts, the electromagnetic field $(\tilde{\bi E}, \tilde{\bi B})$ complies with all the four Maxwell equations, the `unphysicality' of its discontinuity and lack of regular definition on the  spherical surface $r=ct$ notwithstanding.
The delta-function terms in the fields (\ref{finaltildeE}) and (\ref{tildeB}), which are  novel features of  our approach, are a mathematical consequence of the instantaneous jump in the charge's velocity from zero to a nonzero value at the space-time point $(r{=}0,\,t{=}0)$.
   
\section{Coulomb-gauge potentials}
The Lorenz-gauge potentials
 $\tilde{\Phi}_{\rm L}$ and $\tilde{\bi A}_{\rm L}$ of equation\,(\ref{tildePA}) can be transformed  to Coulomb-gauge potentials 
 $\tilde{\Phi}_{\rm C}$ and $\tilde{\bi A}_{\rm C}$ using a gauge function $\chi_{\rm C}({\bi r},t)$ that is defined by the relations
\begin{align}
\frac{\partial\chi_{\rm C}({\bi r},t)}{c\partial t}=\tilde{\Phi}_{\rm L}({\bi r},t)-\tilde{\Phi}_{\rm C}({\bi r},t), \quad
\boldsymbol{\nabla}\chi_{\rm C}({\bi r},t)=\tilde{\bi A}_{\rm C}({\bi r},t)-\tilde{\bi A}_{\rm L}({\bi r},t).
\label{chiC}
\end{align}
The conditions of the Lorenz and Coulomb gauges are   
$\partial\tilde{\Phi}_{\rm L}/c\partial t+\boldsymbol{\nabla}\bdot \tilde{\bi A}_{\rm L}=0$ and $\boldsymbol{\nabla}\bdot\tilde{\bi A}_{\rm C}=0$, respectively, and so  the divergence of the 2nd equality of equation\,(\ref{chiC}) yields  
\begin{align}
\boldsymbol{\nabla}\bdot\boldsymbol{\nabla}\chi_{\rm C}&=\boldsymbol{\nabla}\bdot\tilde{\bi A}_{\rm C}-\boldsymbol{\nabla}\bdot\tilde{\bi A}_{\rm L}
\nonumber \\
&=-\boldsymbol{\nabla}\bdot\tilde{\bi A}_{\rm L}=\partial\tilde{\Phi}_{\rm L}({\bi r},t)/c\partial t.
\label{nablachi}
\end{align}
The gauge function $\chi_{\rm C}$ thus satisfies  Poisson's equation
\begin{align}
\nabla^2 \chi_{\rm C}({\bi r},t)=\frac{\partial\tilde{\Phi}_{\rm L}({\bi r},t)}{c\partial t},
\label{Poisson}
\end{align}
the free-space Green's function for which gives the gauge function  by
\begin{align}
\chi_{\rm C}({\bi r},t)=-\frac{1}{4\pi}\int \frac{\rmd^3r'}{|{\bi r}-{\bi r}'|}\frac{\partial\tilde{\Phi}_{\rm L}({\bi r}',t)}{c\partial t},
\label{solution}
\end{align}
where, using the first equality of equation\,(\ref{tildePA}),
\begin{align}
\frac{\partial\tilde{\Phi}_{\rm L}({\bi r},t)}{c\partial t}=-\beta\frac{\partial\Phi_{\rm L}({\bi r},t)}{\partial x}\,\Theta(ct-r)+\Phi_{\rm L}({\bi r},t)\,\delta(ct-r)-\frac{q}{r}\,\delta(r-ct).
\label{dtildePhidt}
\end{align}
Here, the  factors
$\Theta(ct-r)$ and $\delta(ct-r)$  
make $\partial\tilde{\Phi}({\bi r},t)/c\partial t$ a well-localized function of $\bi r$ at any finite time $t$, ensuring the convergence of the integral representation (\ref{solution}) of the gauge function $\chi_{\rm C}({\bi r},t)$.

It can be checked directly that the first relation of equation\,(\ref{chiC}) is satisfied by the integral representation (\ref{solution}) of $\chi_{\rm C}$. Indeed, we have
\begin{align}
\frac{\partial\chi_{\rm C}({\bi r},t)}{c\partial t}&=-\frac{1}{4\pi}\int \frac{\rmd^3r'}{|{\bi r}-{\bi r}'|}\frac{\partial^2\tilde{\Phi}_{\rm L}({\bi r}',t)}{c^2\partial t^2}\nonumber\\
&=-\frac{1}{4\pi}\int \frac{\rmd^3r'}{|{\bi r}-{\bi r}'|}(\nabla'^2-\Box')\tilde{\Phi}_{\rm L}({\bi r}',t)\nonumber\\
&=-\frac{1}{4\pi}\int \frac{\rmd^3r'}{|{\bi r}-{\bi r}'|}\,[
\nabla'^2\tilde{\Phi}_{\rm L}({\bi r}',t)+4\pi\tilde{\rho}({\bi r}',t)] \nonumber \\
&=-\frac{1}{4\pi}\int\frac{\rmd^3r'}{|{\bi r}-{\bi r}'|}\,\nabla'^2\tilde{\Phi}_{\rm L}({\bi r}',t)
-\tilde{\Phi}_{\rm C}({\bi r},t)\nonumber \\
&=-\frac{1}{4\pi}\int \rmd^3r'\,\tilde{\Phi}_{\rm L}({\bi r}',t)\, \nabla'^2\frac{1}{|{\bi r}-{\bi r}'|}
-\tilde{\Phi}_{\rm C}({\bi r},t)
\nonumber \\
&=\tilde{\Phi}_{\rm L}({\bi r},t)-\tilde{\Phi}_{\rm C}({\bi r},t). 
\end{align}
Here, the integral in the 4th line is transformed by applying integration by parts twice  and, in the last line,  the identity $\nabla'^2|{\bi r}-{\bi r}'|^{-1}=-4\pi\delta({\bi r}-{\bi r}')$ is used. 

We can show also directly that the second relation of equation\,(\ref{chiC}) is satisfied by the integral representation (\ref{solution}) of $\chi_{\rm C}$:
\begin{align}
\boldsymbol{\nabla}\chi_{\rm C}({\bi r},t)&=\frac{1}{4\pi}\int \rmd^3r'\left[\boldsymbol{\nabla}'
\frac{1}{|{\bi r}-{\bi r}'|}\right]\frac{\partial\tilde{\Phi}_{\rm L}({\bi r}',t)}{c\partial t}\nonumber\\
&=\frac{1}{4\pi}\int \rmd^3r'
\frac{1}{|{\bi r}-{\bi r}'|}\boldsymbol{\nabla}'[\boldsymbol{\nabla}'\bdot\tilde{\bi A}_{\rm L}({\bi r}',t)]\nonumber\\
&=\frac{1}{4\pi}\int \rmd^3r'
\frac{1}{|{\bi r}-{\bi r}'|}\,\{\nabla'^2\tilde{\bi A}_{\rm L}({\bi r}',t)+\boldsymbol{\nabla}'\times[\boldsymbol{\nabla}'\times\tilde{\bi A}_{\rm L}({\bi r}',t)]\}\nonumber\\
&=\frac{1}{4\pi}\int \rmd^3r'\tilde{\bi A}_{\rm L}({\bi r}',t)\nabla'^2\frac{1}{|{\bi r}-{\bi r}'|}
+\frac{1}{4\pi}\boldsymbol{\nabla}\times\int\frac{\rmd^3r'}{|{\bi r}-{\bi r}'|}\,\boldsymbol{\nabla}'\times \tilde{\bi A}_{\rm L}({\bi r}',t)\nonumber\\
&=-\tilde{\bi A}_{\rm L}({\bi r},t)+\tilde{\bi A}_{\rm C}({\bi r},t).
\end{align}
Here, in the 2nd line, the integral is transformed by integrating by parts and using the Lorenz-gauge condition; 
in the 4th line, integration by parts is applied twice on the 1st term of the integrand; and,  in last line, the identity 
$\nabla'^2|{\bi r}-{\bi r}'|^{-1}=-4\pi\delta({\bi r}-{\bi r}')$ is used  and the 2nd term in 4th line is recognized as the transverse part of the Lorenz-gauge vector potential $\tilde{\bi A}_{\rm L}$ (integrating by parts, the curl operators can be moved outside the integral) and as such as the Coulomb-gauge vector potential 
$\tilde{\bi A}_{\rm C}$ \cite{VH2}.

Integrating the whole first equality of equation\,(\ref{chiC}) with respect to $t$, we obtain for the gauge function $\chi_{\rm C}$ an expression in terms of a one-dimensional integral:
\begin{align}
\chi_{\rm C}({\bi r},t)=c\int_{t_0}^t \rmd t'\,[\tilde{\Phi}_{\rm L}({\bi r},t')-\tilde{\Phi}_{\rm C}({\bi r},t')]+\chi_0.
\label{chiC2}
\end{align}
The integration term  $\chi_0$ can be shown to be a constant independent of $\bi r$ and $t$ \cite{Jack}, reflecting the fact that a gauge function is defined only to within such an additive  constant, which  we shall omit henceforth. The `instantaneous' Coulomb-gauge scalar potential being easily calculable, the representation (\ref{chiC2}) of 
$\chi_{\rm C}$ is much easier to use for its calculation than representation (\ref{solution}), which is in terms of a three-dimensional integral. With 
\begin{align}
\tilde{\Phi}_{\rm L}({\bi r},t)&=\frac{q\,\Theta(ct-r)}{\sqrt{(x-vt)^2+(y^2+z^2)/\gamma^2}} +\frac{q}{r}\,
\Theta(r{-}ct),
\label{tildePL} \\  
\tilde{\Phi}_{\rm C}({\bi r},t)&=\frac{q\,\Theta(t)}{\sqrt{(x-vt)^2+y^2+z^2}}+\frac{q}{r}\,\Theta(-t)
\label{tildePC}
\end{align}
the time integration in (\ref{chiC2}) can be performed in closed form, yielding for the gauge function 
$\chi_{\rm C}$ an expression that is only a little more involved than that for a uniformly moving charge:\footnote{[2], equation\,(13); that expression for $\chi_{\rm C}$ can be simplified by replacing $x-x_0$ with $\gamma(x-vt)$, utilizing the fact that the gauge function is defined only to within a term independent of $\bi r$ and $t$.} 
\begin{align}
\chi_{\rm C}({\bi r},t)&=q\left[{\rm arsinh}\frac{\gamma(x-\beta r)}{\sqrt{y^2+z^2}}-{\rm arsinh}\frac{\gamma(x-vt)}{\sqrt{y^2+z^2}}\right]\frac{\Theta(ct-r)}{\beta}\nonumber\\
&\quad+q\left[{\rm arsinh}\frac{x-vt}{\sqrt{y^2+z^2}}-{\rm arsinh}\frac{x}{\sqrt{y^2+z^2}}\right]
\frac{\Theta(t)}{\beta}\nonumber\\
&\quad+\frac{q}{r}[(r-ct)\Theta(ct-r)+ct\Theta(t)].
\label{closedchiC}
\end{align}
The differentiation of (\ref{closedchiC}) with respect to time and the use of the identity 
$f(x)\delta({x-}x_0)=f(x_0)\delta(x{-}x_0)$ in the delta-function terms arising from the derivatives of the Heaviside step functions confirm that the gauge function (\ref{closedchiC}) satisfies the first defining relation of equation\,(\ref{chiC}). It can be shown also using the delta-function identities that include $[f(x)-f(x_0)]\delta'(x-x_0)=-f'(x_0)\delta(x-x_0)$ \cite{Kanwal} that the Laplacian of (\ref{closedchiC})
equals expression (\ref{dtildePhidt}), confirming that the gauge function (\ref{closedchiC}) satisfies Poisson's equation (\ref{Poisson}).

We can now calculate the Coulomb-gauge vector potential $\tilde{\bi A}_{\rm C}$ using the second defining relation of equation\,(\ref{chiC}) with the gauge function (\ref{closedchiC}) and the Lorenz-gauge vector potential
of equation\,(\ref{tildePA}). The resulting components of $\tilde{\bi A}_{\rm C}$ are:
\begin{align}
\tilde{A}_{{\rm C}x}({\bi r},t)&=\frac{\partial \chi_{\rm C}}{\partial x} +\tilde{A}_{{\rm L}x}\nonumber \\
&=q\left[\frac{r-\beta x}{r\sqrt{(x{-}\beta r)^2+s^2/\gamma^2}} -\frac{1-\beta^2}{\sqrt{(x{-}vt)^2+s^2/\gamma^2}}\right]\frac{\Theta(c t{ -}r)}{\beta}\nonumber\\
&\quad +q\left[\frac{1}{\sqrt{(x{-}vt)^2+s^2}}-\frac{1}{r}\right]\frac{\Theta(t)}{\beta}
+q\frac{c t x}{r^3}\,[\Theta(ct-r)-\Theta(t)], \label{tildeACx}\\
\tilde{A}_{{\rm C}y}({\bi r},t)&=\frac{\partial \chi_{\rm C}}{\partial y}
=q\left[\frac{(x-vt)y}{\sqrt{(x{-}vt)^2+s^2/\gamma^2}}-\frac{(r-\beta x)xy}{r 
\sqrt{(x{-}\beta r)^2+s^2/\gamma^2}}\right]\frac{\Theta(ct{-}r)}{\beta s^2}\nonumber \\
&\quad-q\left[\frac{(x-vt)y}{\sqrt{(x{-}vt)^2+s^2}}-\frac{xy}{ r}\right]
\frac{\Theta(t)}{\beta s^2}+q\frac{cty}{r^3}[\Theta(ct-r)-\Theta(t)], \label{tildeACy} \\
\tilde{A}_{{\rm C}z}({\bi r},t)&=\frac{\partial \chi_{\rm C}}{\partial z}
=q\left[\frac{(x-vt)z}{\sqrt{(x{-}vt)^2+s^2/\gamma^2}}-\frac{(r-\beta x)xz}{ r 
\sqrt{(x{-}\beta r)^2+s^2/\gamma^2}}\right]\frac{\Theta(ct{-}r)}{\beta s^2}\nonumber \\
&\quad-q\left[\frac{(x-vt)z}{\sqrt{(x{-}vt)^2+s^2}}-\frac{xz}{r}\right]
\frac{\Theta(t)}{\beta s^2}+q\frac{ctz}{r^3}[\Theta(ct-r)-\Theta(t)], \label{tildeACz}
\end{align}
where $s=(y^2+z^2)^{1/2}$. The use of  this vector potential and of the Coulomb-gauge scalar potential (\ref{tildePC}) in
$\tilde{\bi E}=-\boldsymbol{\nabla}\tilde{\Phi}_{\rm C}-\partial \tilde{\bi A}_{\rm C}/c\partial t$ and 
$\tilde{\bi B}= \boldsymbol{\nabla}\times \tilde{\bi A}_{\rm C}$
results in electric and magnetic fields that are the same as those of equations\,(\ref{finaltildeE}) and (\ref{tildeB}), obtained using the Lorenz-gauge potentials,
confirming the correctness of expressions (\ref{tildeACx}--\ref{tildeACz}). These calculations are cumbersome but straightforward;  
all but one of the delta-function terms due to the derivatives of the Heaviside-step-function factors vanish on account of the identity $f(x)\delta(x-x_0)=f(x_0)\delta(x-x_0)$.

\section{Concluding remarks}
We found the Lorenz-gauge and Coulomb-gauge potentials of a point charge that is set suddenly into uniform motion. The Coulomb-gauge vector potential  was obtained using the pertinent gauge function of the transformation between those two gauges, calculated in closed form. The electric field of the charge calculated using the potentials agrees with that found by Purcell \cite{Pur} by just employing imaginatively the fact that electromagnetic effects propagate at the speed of light. 

The electromagnetic field of the charge complies with all the four Maxwell  equations, despite the `unphysical' instantaneous jump of the charge's velocity from zero to a nonzero value. In any case, it can be assumed that the charge attains its constant  velocity in an initial time interval of a finite, but arbitrarily short duration $\tau$ by application of an external force of sufficiently high magnitude. In the limit $\tau\rightarrow 0$, the regular electromagnetic field that would connect the field in the regions interior and exterior to the sphere of radius $r=ct$ is replaced by the delta-function terms of the fields (\ref{finaltildeE}) and (\ref{tildeB}).


\section*{Acknowledgments}
We thank David Griffiths, Kirk McDonald and Dragan Red\v{z}i{\'c} for useful comments on drafts. The anonymous referees are thanked for constructive suggestions. 
A recent eprint of V Onoochin \cite{On2}, in which the case of a `charge that starts' is used, has provided a stimulus for doing it properly. 

\section*{Data availability statement}
No new data were analysed or created in this study.

\section*{References}

\end{document}